\documentclass[12pt]{article}

\usepackage[utf8]{inputenc}

\usepackage{graphicx}
\usepackage{float}
\usepackage{subfig}
\usepackage{amsmath}
\usepackage{natbib}
\usepackage{color}
\usepackage{multirow}
\usepackage{longtable}
\usepackage{rotating}
\usepackage{verbatim}
\usepackage{booktabs}
\usepackage{algorithm}
\usepackage{algorithmic}
\usepackage{rotating}
\usepackage{natbib}
\usepackage[OT1]{fontenc}
\usepackage{hyperref}

\usepackage[margin=1in]{geometry}

\usepackage{bm}
\usepackage{amssymb}

\newcommand{\dd}{\mathrm{d}}

\begin{document}



\title{A Bayesian Marked Spatial Point Processes Model for Basketball Shot Chart}
\author{Jieying Jiao,~~Guanyu Hu,~~Jun Yan}


\maketitle

\abstract{
The success rate of a basketball shot may be higher at
locations where a player makes more shots. For a marked spatial
point process, this means that the mark and the intensity are
associated. We propose a Bayesian joint model for the mark and the
intensity of marked point processes, where the intensity is
incorporated in the mark model as a covariate. Inferences are
done with a Markov chain Monte Carlo algorithm. Two Bayesian
model comparison criteria, the Deviance Information Criterion and the
Logarithm of the Pseudo-Marginal Likelihood, were
used to assess the model.
The performances of the proposed methods were
examined in extensive simulation studies. The proposed methods were
applied to the shot charts of four players (Curry, Harden, Durant,
and James) in the 2017--2018 regular season of the National Basketball
Association to analyze their
shot intensity in the field and the field goal percentage in detail.
Application to the top 50 most frequent shooters in the season
suggests that the field goal percentage and the shot intensity are
positively associated for a majority of the players. The fitted
parameters were used as inputs in a secondary analysis to cluster the
players into different groups.
}

\section{Introduction}\label{sec:intro}

Shot charts are important summaries for basketball players.
A shot chart is a spatial representation of the location and the
result of each shot attempt by one player.
Good defense strategies depend on good understandings of
the offense players' tendencies to shoot and abilities to score.
\citet{reich2006spatial}
proposed hierarchical spatial models with spatially-varying covariates
for shot attempt frequencies over a grid on the court and for shot
success with shot locations fixed. Spatial point processes are commonly used
to model random locations
\citep[e.g.,][]{cressie2015statistics, diggle2013statistical}.
\citet{miller2014factorized}
used a low dimensional representation of related point processes to
analyze shot attempt locations. \citet{franks2015characterizing}
combined spatial and spatio-temporal processes, matrix factorization
techniques, and hierarchical regression models to analyze defensive
skill. Many parametric models
for spatial point process have been proposed in the literature,
such as the Poisson process \citep{geyer1998likelihood},
the Gibbs process \citep{moller2003statistical},
and the log Gaussian Cox process (LGCP) \citep{moller1998log}.
When each point in a point process is companied with a random variable or vector
known as mark, the resulting process is a marked point process
\citep[e.g.,][Ch.~8]{banerjee2014hierarchical}.
A shot chart can be modeled by a spatial marked point process with a binary
mark showing the shot results.

The frequency of successful shots may be higher at locations where a
player makes more shot attempts. This positive association
is expected from different angles. More frequent shots suggests higher
competence level and, hence, higher shooting accuracy. 
Higher accuracy also encourages more shooting since it means higher reward. In
behavioral science, the matching law states that individuals will allocate their
behavior according to the relative rates of reinforcement available for each option
\citep{baum1974on, staddon1978theory}.
It predicts higher proportion of 3-point shots taken relative to all
shots to be associated with higher proportion of 3-point shots
scored relative to all shots scored \citep{vollmer2000application,
  alferink2009generality}. The association might be more obvious for
players who get fewer minutes and may be more selective to
``prove their worth''. It might be less so for players who have possession more
often and may be less selective.
Team strategies could affect the association in two opposite directions. Players
with high three-point shot accuracy are more likely to be arranged in areas beyond the
three-point line. This is in favor of the positive association.
On the other hand, optimal shot selection strategy requires all shot locations
to have the same marginal shot efficiency for the whole team \citep{skinner2015optimal},
which may not be consistent with shot selections for individual players.
A quantitative measure of the association for a player will be helpful for
understanding the player's performance and suggesting directions for
improvement at both player level and team strategies level.

We consider
marked spatial point processes where the mark distributions depend on the
point pattern. There are two approaches to model this dependence.
Location dependent models \citep{mrkvivcka2011spatial}
are observation driven, where the observed point pattern is
incorporated into characterizing the spatially varying distribution of
the mark. Intensity dependent models \citep{ho2008modelling} are
parameter driven, where the intensity instead of the observed point
pattern characterizes the distribution of the
mark at each point in the spatial domain.
For basketball shot charts, no work has jointly modeled
the intensity of the shot attempts and the results of the attempts.

The contribution of this paper is two-fold.
First, we propose a Bayesian joint model for marked spatial point
processes to study the association between shot intensity and shot
accuracy. In particular, we use a non-homogeneous
Poisson point process to model the spatial pattern of the shot attempts and
incorporate the shot intensity as a covariate in the model of shot accuracy.
Inferences are made with Markov chain Monte Carlo (MCMC).
The deviance information criterion (DIC) and the 
logarithm of the pseudo-marginal likelihood (LPML) are used to
assess the fitness of our proposed model.
Our second contribution is the analyses of four representative
players and the top 50 most frequent shooters in the 2017--2018
regular season of the National Basketball Association (NBA).
The shot intensity of each player is captured by a set of intensity
basis constructed from historical data which represents different shot
types such as long 2-pointers and corner threes, among others
\citep{miller2014factorized}. For a majority
(about 80\%) of the these players, the results support a significant positive association
between the shot accuracy and shot intensity. The fitted coefficients
are then used as input for a clustering analysis to group the top 50 most
frequent shooters in the season, which provides insights
for game strategies and training management.

The rest of the paper is organized as follows.
In Section~\ref{sec:data}, the shot charts of selected players from
the 2017--2018 NBA regular season, along with research questions
that such data can help answer, are introduced.
In Section~\ref{sec:method}, we develop the Bayesian joint model of
marked point process. Details of the Bayesian computation are presented in
Section~\ref{sec:bayescomp}, including the MCMC algorithm and the
two model selection criteria. Extensive simulation studies are
summarized in Section~\ref{sec:simu} to investigate empirical
performance of the proposed methods. Applications of the proposed
methods to four NBA players are reported in
Section~\ref{sec:app}. Section~\ref{sec:disc} concludes with a
discussion.

\section{Shot Charts of NBA Players}\label{sec:data}

We focus on the 2017--2018 regular NBA season here.
The website \url{NBAsavant.com} provides a convenient tool to search
for shot data of NBA players, and the original data are a consolidation between
the NBA statistics (\url{https://stats.nba.com}) and ESPN's
shot tracking (\url{https://shottracker.com}). For each player, the
available data contains information about each of his shots
in this season including game date, opponent team, game period when
the shot was made (four quarters and a fifth period representing extra
time), minutes and seconds left to the end of that period, success
indicator or mark (0 for missed and 1 for made), shot type (2-point or
3-point shot), shot distance, and shot location coordinates, among
others. Euclidean shot distances were rounded to foot.

\begin{figure}[tbp]
\centering
\includegraphics[width = \textwidth]{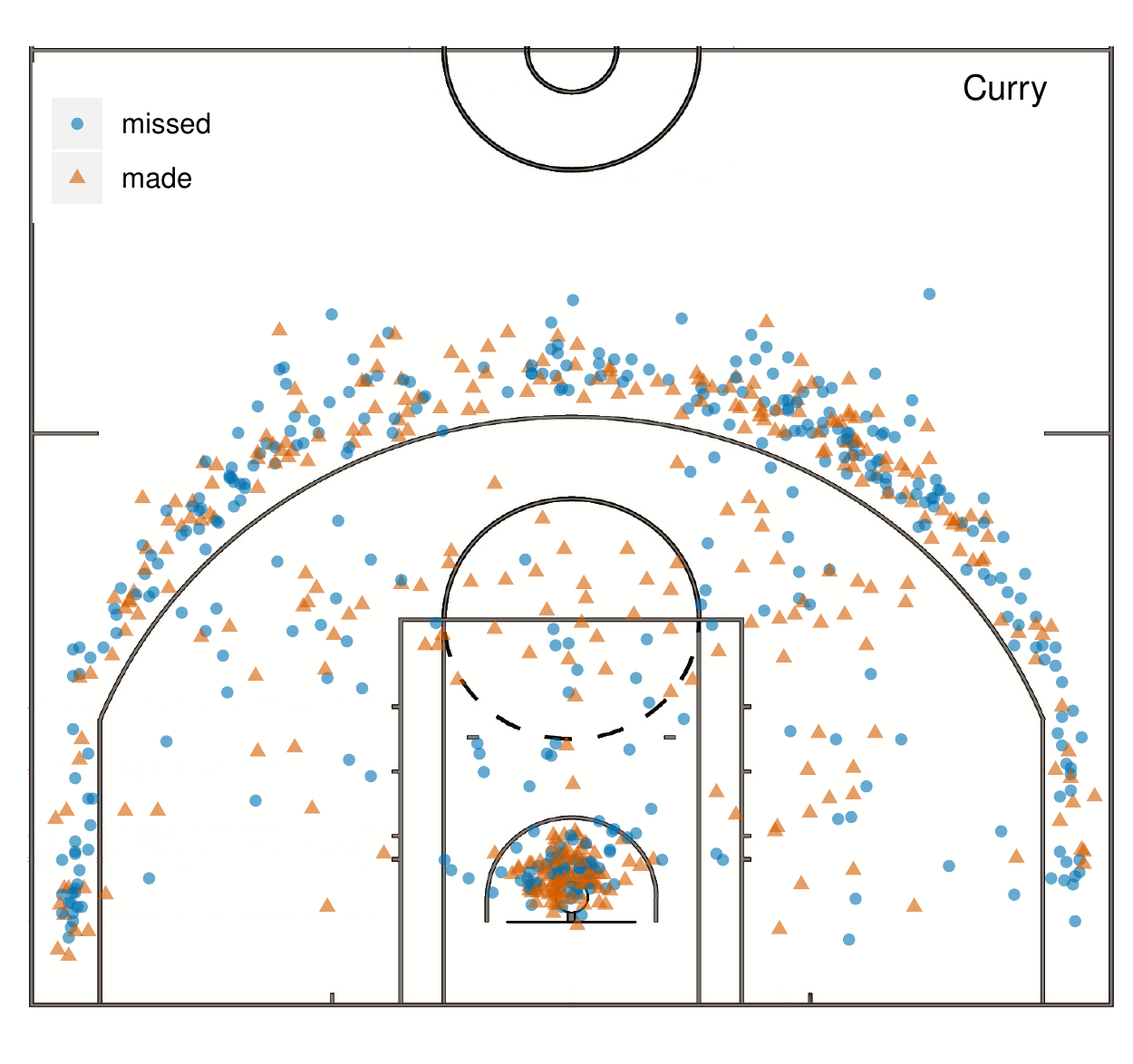}
\caption{Shot charts of Curry in the
    2017--2018 regular NBA season.}
  \label{fig:EDA}
\end{figure}

We chose four famous players
with quite different styles: Stephen Curry, Kevin Durant, James Harden
and LeBron James. Figure~\ref{fig:EDA} shows Curry's shot locations with
the shot success indicators as a demonstration. The total number of shots was in
the range of 740 (Curry) to 1409 (James). Curry has the highest proportion of
3-point shots (57\%) while James made the highest proportion of
2-point shots (75\%). The field goal percentage ranged from 45\%
(Harden) to 52\% (James). As shown in Figure~\ref{fig:EDA}, most of
the shots were made close to the rim or out of but close to the
3-point line. This is expected since shorter distance should give
higher shot accuracy for either 2-point or 3-point shots.

The shot chart of each player can be modeled by a marked point
process that captures the dependence between the binary mark and the
intensity of the shots. Through analyses of the selected NBA players, we
address the following questions: How to characterize the shot pattern of
individual players? What are the factors, such as shot location, time remaining,
period of the game, and the level of the opponent, that may affect the shot
accuracy?  Is there a positive association between shot accuracy and shot
intensity of some players? How often is the positive association seen among
the most frequent shooters? Is this positive association different between
2-point versus 3-point shots? Can the players be grouped by their shooting
styles? These questions may not be completely answered, but even partial answers
would shed lights on understanding the game and the players for better
game strategies and training management.

\section{Bayesian Marked Spatial Point Process Model}
\label{sec:method}
The observed shot chart of a player can be represented by
$(\mathbf{S}, \mathbf{M})$, where $\mathbf{S}$ is the collection of
the locations of shot attempts ($x$ and $y$ coordinates) and
$\mathbf{M}$ is the vector of the
corresponding marks (1~means success and 0~means failure).
Assuming that $N$ shots were observed, we have
$\mathbf{S} = (\bm{s}_1, \bm{s}_2, \dots, \bm{s}_N)$ and
$\mathbf{M} = (m(\bm{s}_1), m(\bm{s}_2), \dots, m(\bm{s}_N))$.

\subsection{Marked Spatial Point Process}
We propose to model $(\mathbf{S}, \mathbf{M})$ by a marked spatial
point process. The shot locations~$\mathbf{S}$ are modeled by a non-homogeneous
Poisson point process \citep[e.g.,][]{diggle2013statistical}.
Let $\mathcal{B} \subset \mathbb{R}^2$ be a subset of the half
basketball court on which we are interested in modeling the shot
intensity.  A Poisson point process is defined such that
$N(A) = \sum_{i = 1}^N 1(\bm{s}_i \in A)$ for any
$A \subset \mathcal{B}$ follows a Poisson distribution with mean
$\lambda(A) = \int_A \lambda(\bm{s}) \dd \bm{s}$, where
$\lambda(\cdot)$ defines an intensity function of the process.
The likelihood of the observed locations $\mathbf{S}$ is
\begin{align*}
\prod_{i=1}^N \lambda(\bm{s}_i)
            \exp\left(-\int_{\mathcal{B}} \lambda(\bm{s}) \dd \bm{s} \right).
\end{align*}
Covariates can be incorporated into the intensity by setting
\begin{align}
  \lambda(\bm{s}_i)=\lambda_0\exp\big(\mathbf{X}^\top(\bm{s}_i)\bm{\beta}\big),
  \label{eq:IntenReg}
\end{align}
where $\lambda_0$ is a baseline intensity, $\mathbf{X}(\bm{s}_i)$ is a
$p \times 1$ spatially varying covariate vector, and
$\bm{\beta}$ is the corresponding coefficient vector.

Next we consider modeling the success indicator (mark).
It is natural to suspect that the success rate of shot attempts is
higher at locations with higher shot intensity, suggesting
an intensity dependent mark model. In particular, the success
indicator is modeled by a logistic regression
\begin{align}
\begin{split}
m(\bm{s}_i) \mid \mathbf{Z}(\bm{s_i}) &\sim
\mathrm{Bernoulli}\big(\theta(\bm{s}_i)\big),\\
  \mathrm{logit}\big(\theta(\bm{s}_i)\big) &= \xi\lambda(\bm{s}_i) +
\mathbf{Z}^\top(\bm{s}_i)\bm{\alpha},
\end{split}
\label{eq:logistic}
\end{align}
where $\lambda(\bm{s}_i)$ is the intensity defined
in~\eqref{eq:IntenReg} with a scalar coefficient $\xi$,
$\mathbf{Z}(\bm{s}_i)$ is a $q \times 1$
covariate vector evaluated at $i$-th data point ($\mathbf{Z}$ does
not need to be spatial, like period covariates), and $\bm{\alpha}$ is a $q\times 1$
vector of coefficient.

With $\bm{\Theta} = (\lambda_0, \bm{\beta}, \xi, \bm{\alpha})$, the
joint likelihood for the observed marked spatial point process
$(\mathbf{S}, \mathbf{M})$ is
\begin{align}
  L(\bm{\Theta}\mid \mathbf{S}, \mathbf{M})
  \propto &\prod_{i=1}^N \theta(\bm{s}_i)^{m(\bm{s}_i)}
               \left(1 - \theta(\bm{s}_i)\right)^{1 - m(\bm{s}_i)}\notag\\
          &\times \left(\prod_{i=1}^N \lambda(\bm{s}_i)\right)
            \exp\left(-\int_{\mathcal{B}} \lambda(\bm{s}) \dd \bm{s} \right).
  \label{eq:jointll}
\end{align}

\subsection{Prior Specification}
Vague priors are specified for model parameters.
For $\lambda_0$, the gamma distribution is conjugate
prior \citep[e.g.,][]{leininger2017bayesian}. For $\bm{\beta}$, $\xi$,
or $\bm{\alpha}$, there is no 
conjugate prior and we specify a vague, independent normal
prior. In summary, we have
\begin{align}
  \begin{split}
    \lambda_0 &\sim \mathrm{G}(a, b),\\
    \bm{\beta} &\sim \mathrm{MVN}(\bm{0},\sigma^2_{\beta}\mathbf{I}_p),\\
    \xi &\sim \mathrm{N}(0, \sigma^2_{\xi}),\\
    \bm{\alpha} &\sim \mathrm{MVN}(\bm{0}, \sigma^2_{\alpha}\mathbf{I}_{q}),
  \end{split}
\label{eq:mppprior}
\end{align}
where $\mathrm{G}(a, b)$ represents a Gamma distribution with
shape~$a$ and rate~$b$, respectively, $\mathrm{MVN}(\bm{0},
\bm{\Sigma})$ is a multivariate normal distribution with mean vector
$\bm{0}$ and variance matrix $\bm{\Sigma}$, $(a, b, \sigma^2_\beta,
\sigma^2_\xi, \sigma^2_\alpha)$ are
hyper-parameters to be specified, and $\mathbf{I}_k$ is the
$k$-dimensional identity matrix.

\section{Bayesian Computation}
\label{sec:bayescomp}

\subsection{The MCMC Sampling Schemes}
The posterior distribution of $\bm{\Theta}$ is
\begin{align}
  \pi(\bm{\Theta}|\mathbf{S},\mathbf{M}) &\propto
                              L(\bm{\Theta} \mid \mathbf{S}, \mathbf{M})
                                    \pi(\bm{\Theta}),
  \label{eq:posterior}
\end{align}
where
$\pi(\bm{\Theta}) = \pi(\lambda_0)\pi(\bm{\beta})\pi(\xi)\pi(\bm{\alpha})$
is the joint prior density as specified in~\eqref{eq:mppprior}. In
practice, we used vague priors with 
hyper-parameters $\sigma^2_\beta = \sigma^2_\xi = \sigma^2_\alpha = 100$ and
$a = b = 0.01$ in~\eqref{eq:mppprior}.

To sample from the posterior distribution of $\bm{\Theta}$ in
\eqref{eq:posterior}, an Metropolis--Hasting within Gibbs
algorithm is facilitated by R package \textsf{nimble}
\citep{de2017programming}. The loglikelihood function of the joint
model used in the MCMC iteration is directly defined using the
\texttt{RW\_llFunction()} sampler.
The integration in the likelihood function~\eqref{eq:jointll}
does not have a closed-form. It needs to be computed with a
Riemann approximation by partitioning $\mathcal{B}$ into a grid with a
sufficiently fine resolution. Within each grid box, the integrand
$\lambda(\bm{s})$ is approximated by a constant. Then the integration of
$\lambda(\bm{s})$ becomes a summation over all of the grid boxes.

\subsection{Bayesian Model Comparison}
To assess whether the intensity term is necessary in the mark
model~\eqref{eq:logistic}, model comparison criteria is needed.
Within the Bayesian framework, DIC
\citep{spiegelhalter2002bayesian} and LPML
\citep[LPML;][]{geisser1979predictive, gelfand1994bayesian}
are two well-known Bayesian criteria for model comparison.
Using the method of \citet{zhang2017bayesian}, each criterion for the proposed
joint model can be decomposed into one for the intensity model and one for the
mark model conditioning on the point pattern for more insight on the model
comparison.

The DIC for the joint model is
\begin{align}
  \begin{split}
    \text{DIC}
   &= \text{Dev}(\bar{\bm{\Theta}} \mid
     \textbf{S}, \textbf{M}) + 2 p_D,\\
  p_D &= \overline{\text{Dev}}(\bm{\Theta} \mid
    \textbf{S}, \textbf{M}) - \text{Dev}(\bar{\bm{\Theta}}\mid
    \textbf{S}, \textbf{M}),
  \end{split}
\label{eq:DIC}
\end{align}
where the deviance $\text{Dev}$ is the negated loglikelihood function
in Equation~\eqref{eq:jointll},
$\overline{\text{Dev}}$ is the mean of the deviance evaluated at each
posterior draw of the parameters, $\bar{\bm{\Theta}}$ is the posterior mean
of $\bm{\Theta}$, and $p_D$ is known as the effective number of parameters.
For the intensity model and the conditional mark model, the DIC can be computed
with deviance, respectively,
\begin{align}
  \begin{split}
  \text{Dev}_{\text{intensity}}(\lambda_0, \bm{\beta}\mid \textbf{S}) &= -2 
  \left(\sum_{i = 1}^N\log \lambda(\bm{s}_i)- \int_{\mathcal{B}}\lambda(\bm{s}) \dd \bm{s} \right),\\
  \text{Dev}_{\text{mark}}(\bm{\lambda},\bm{\alpha}, \xi \mid \textbf{M}, \textbf{S})&=-2
    \sum_{i = 1}^N\log f (m(\bm{s}_i) \mid \textbf{S};
    \lambda(\bm{s}_i),\bm{\alpha}, \xi, \textbf{Z}(\bm{s}_i)),
  \end{split}
  \label{eq:dev}
\end{align}
where
$\bm{\lambda} = (\lambda(\bm{s}_1), \lambda(\bm{s}_2), \dots,\lambda(\bm{s}_N))$,
and $f (m(\bm{s}_i) \mid \lambda(\bm{s}_i),\bm{\alpha}, \xi,
\textbf{Z}(\bm{s}_i))$ is the conditional probability mass function of
$m(\bm{s}_i)$ given
$\left(\lambda(\bm{s}_i),\bm{\alpha}, \xi, \textbf{Z}(\bm{s}_i)\right)$.
Clearly, the DIC for joint model is the summation of the DIC for the
intensity model and the DIC for the conditional mark model.
Models with smaller DIC are better models.

Calculation of the LPML for point process models is challenging because the usual
conditional predictive ordinate (CPO) based on the leaving-one-out assessment is
not applicable where the number of points $N$ is random. \citet{hu2019new}
recently suggested a Monte Carlo method to approximate the LPML for the
intensity model as
\begin{align}
  \widehat{\text{LPML}}_{\text{intensity}} = \sum_{i = 1}^N \log \tilde{\lambda}(\bm{s}_i) -
  \int_{\mathcal{B}} \bar{\lambda}(\bm{s}) \dd \bm{s},
  \label{eq:LPML_intensity}
\end{align}
where $\tilde{\lambda}(\bm{s}_i) =
(\frac{1}{K}\sum_{k = 1}^K\lambda^{(k)}(\bm{s}_i)^{-1})^{-1}$, $\bar{\lambda}(\bm{s}) =
\frac{1}{K}\sum_{k = 1}^K \lambda^{(k)}(\bm{s})$, and
$\{\lambda^{(k)}(\bm{s}_i): k = 1, 2, \dots, K\}$
is a posterior sample of size $K$ of the parameters from the MCMC.
The LPML for the conditional mark model can be calculated as usual
\citep[Ch.~10]{chen2000monte}. For the $i$-th data point, define
\begin{align*}
\widehat{\text{CPO}}_i^{-1} = \frac{1}{K}\sum_{b=1}^{K}
\frac{1}{f \big(m(\bm{s}_i) \mid \lambda^{(k)}(\bm{s}_i),\bm{\alpha}^{(k)}, \xi^{(k)},
  \mathbf{Z}(\bm{s}_i)\big)}, 
\end{align*}
where $\{\bm{\alpha}^{(k)}, \xi^{(k)}: k = 1, 2, \dots, K\}$ is a posterior
sample of size $K$ of the parameters from the MCMC. Then the LPML on mark model
is
\begin{align}
\widehat{\text{LPML}}_{\text{mark}} =
  \sum_{i=1}^{N}\text{log}(\widehat{\text{CPO}}_i).
  \label{eq:LPML_mark}
\end{align}
The LPML for the joint model is then calculated as the sum
of~\eqref{eq:LPML_intensity} and~\eqref{eq:LPML_mark}.
Models with higher LPML are better models.

\section{Simulation Studies}\label{sec:simu}
To investigate the performance of the estimation, we generated data
from a non-homogeneous Poisson point process defined on a square
$\mathcal{B} =  [-1,1]\times[-1, 1]$
with intensity 
$\lambda(\bm{s}_i) = 100 \lambda_0 \exp (\beta_1 x_i + \beta_2 y_i)$,
where $\bm{s}_i = (x_i, y_i) \in \mathcal{B}$ is the location for
every data point. For each $\bm{s}_i$, $i = 1, \ldots, N$, the mark
$m(\bm{s}_i)$ follows a logistic model with two covariates in addition to
$\lambda$ and intercept:
\begin{align}
  \label{eq:simuset}
  \begin{split}
    m(\bm{s}_i) &\sim \mathrm{Bern}(p_i),\\
    \mathrm{logit}(p_i) &= \xi\lambda(\bm{s}_i) + \alpha_0 + \alpha_1 Z_{1i} +
    \alpha_2 Z_{2i}.
  \end{split}
\end{align}

The parameters of the model were designed to give point counts
that are comparable to the basketball shot chart data.
We fixed $(\beta_1, \beta_2) = (2, 1)$, $\xi = 0.5$, 
$\alpha_0 = 0.5$, and $\alpha_2 = 1$.
Three levels of $\alpha_1$ were considered,
$\alpha_1 \in \{0.8, 1, 2\}$,
in order to compare the performance of the estimation procedure under
different magnitudes of the coefficients in the mark model.
Two levels of $\lambda_0$ were considered, $\lambda_0\in \{0.5, 1\}$,
which controls the mean of the number of points on $\mathcal{B}$.
It is easy to integrate in this case the intensity function over $\mathcal{B}$
to get the average number of points being 850 and 1700, respectively, for
$\lambda_0 = 0.5$ and $1$. The numbers are approximately in the range
of the NBA basketball shot charts in Section~\ref{sec:data}. In the
mark model, covariate $Z_1$ was generated from the standard normal
distribution; two types of $Z_2$ were considered, standard normal
distribution or Bernoulli with rate~0.5. The resulting range of the 
Bernoulli rate of the marks was within $(0.55, 0.78)$ for all the
scenarios.

For each setting, 200 data sets were generated. R package
\textsf{spatstat} \citep{baddeley2005spatstat} was used to generate
the Poisson point process data with the given intensity function.
The priors for the model parameters were set to
be~\eqref{eq:mppprior} with the hyper-parameters
$\sigma^2_\beta = \sigma^2_\xi = \sigma^2_\alpha=100$ and $a = b =
0.01$. The grid used to calculate the integration in likelihood
function had resolution $100\times 100$. For each data set, a MCMC was
run for 20,000 iterations with the first 10,000 treated as burn-in
period. For each parameter, the posterior mean was used as the point
estimate and the 95\% credible interval was constructed with the
$2.5\%$ lower and upper quantiles of the posterior sample.

Table~\ref{tab:EstSimu_norm}--\ref{tab:EstSimu_binary} in Appendix summarize
the simulation results for the scenarios of standard normal $Z_2$ and
Bernoulli $Z_2$, respectively. The empirical bias for all the settings
are close to zero. The average posterior standard deviation from the
200~replicates is very close to the empirical standard deviation of
the 200 point estimates for all the parameters, suggesting that the
uncertainty of the estimator are estimated well. Consequently, the
empirical coverage rates of the credible intervals are close to the nominal
level~$0.95$. As $\alpha_1$ increases, the variation increases in the
mark parameter estimates but
does not change in the intensity parameter estimates.
As $\lambda_0$ increases, the variations of the estimates for both
intensity and mark parameters get lower. Between the continuous and
binary cases of $Z_2$, the variation in the estimates is higher in the
latter case, especially for the coefficient of $Z_2$.

\section{NBA Players Shot Chart Analysis}\label{sec:app}

\subsection{Covariates Construction}
To capture the shot styles of individual players in their shot intensity model,
we follow \citet{miller2014factorized} to construct basis covariates that are
interpreted as archetypal intensities or ``shot types'' used by the players. The
focus is on the 35 ft by 50 ft rectangle on the side of the backboard in the
offensive half court. The origin of the Cartesian coordinates $(x, y)$ is
replaced at the bottom left corner so that $x \in [0, 50]$ and $y \in [0, 35]$.
The rectangle was evenly partitioned into $50 \times 35$ grid boxes of 1~ft by
1~ft. Our bases construction is slightly different from that of
\citet{miller2014factorized} in the preparation for the Nonnegative matrix
factorization (NMF). First, we used a kernel estimation instead of an LGCP model
to estimate the $50 \times 35$ intensity matrix of each individual players,
which is easier to compute and more accurate in the sense of intensity fitting
accuracy. Second, we used historical data instead of the current season data.
In particular, a kernel estimate of the $50\times 35$ intensity matrix for each
of the 407 players in the previous season (2016--2017) who had made over 50
shots was used as input for the NMF. As in \citet{miller2014factorized}, we
obtained 10 bases using R package \textsf{NMF} \citep{Renaud2010flexible}.

\begin{figure}[tbp]
\centering
\includegraphics[width = \textwidth]{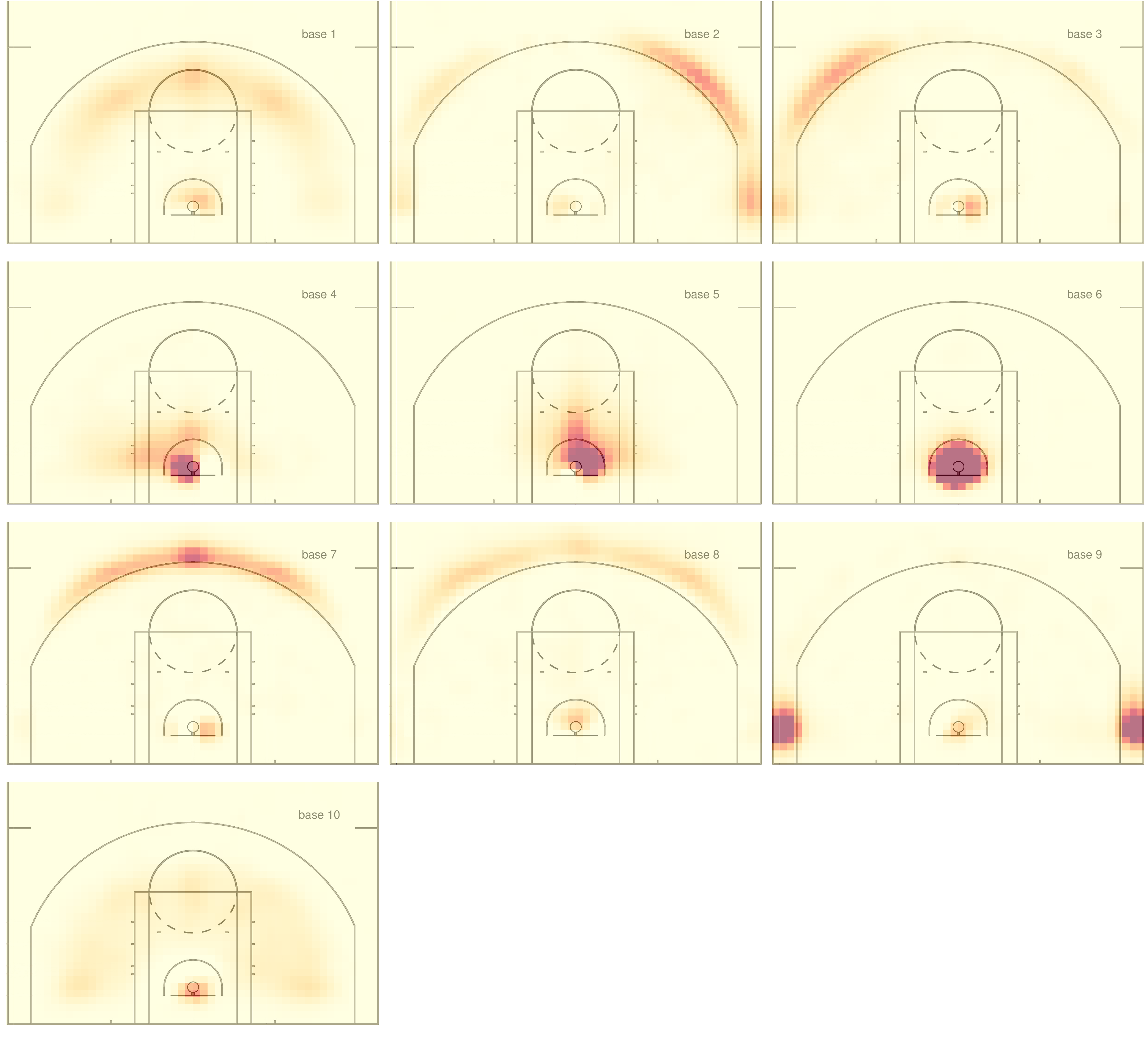}
\caption{Intensity matrix bases heat plots.}
\label{fig:bases_plot}
\end{figure}

Figure~\ref{fig:bases_plot} displays the 10 nonnegative matrix bases that can be
used as covariates for the intensity matrix fitting. They are similar to those
in the literature \citep{miller2014factorized, franks2015characterizing}. Each
basis is nicely interpreted as a certain shot type. For example, basis~1
is long 2-points, bases~2--3 are left/right wing threes, bases~4--5 are
left/right/center restricted area 2-points, basis~7 is top of key threes,
basis~8 is center threes,  basis~9 is corner threes, and basis~10 is mid-range
twos. When used as covariates in modeling individual shot
intensity, their coefficients characterize the shooting style of each player.

The influence of intensity on shot accuracy might be different for different
shot type. Players' shot selection may be biased towards three point shot for
higher reward \citep{alferink2009generality, skinner2015optimal}.
This can result in higher intensity for 3-point shot at locations with not high
accuracy. To capture this tendency, an interaction term between the intensity
and the shot type is introduced to the mark model. In addition to intensity and
interaction term between intensity and shot type, other covariates in the mark
model include distance to the basket and non-spatial covariates such as
seconds left to the end of the period, dummy variables for five different
periods with first period as reference, and the indicator of opponent made to
the playoff in the last season.

\subsection{Model Comparison}
The joint model~\eqref{eq:IntenReg}--\eqref{eq:logistic} was fitted for each
player with the hyper-parameters in~\eqref{eq:mppprior} set as
$\sigma^2_\beta = \sigma^2_\xi = \sigma^2_\alpha=100$
and $a = b = 0.01$. The numerical integration in evaluating the joint
log-likelihood~\eqref{eq:jointll} was based on the same $50 \times 35$ grid
as that used in constructing the basis shot styles from NMF.
To check the importance of intensity as covariate
in the mark component, we also fitted the model with the restriction
$\xi = 0$. For each model fitting, 60,000 MCMC iterations were run. The first
20,000 were discarded as the burn-in period and the rest were thinned by 10,
which led to an MCMC sample of size 4,000. The trace plots of the MCMC were
checked and the convergence of all the parameters were confirmed. The reported
results were obtained from a second run after insignificant covariates were
removed to avoid possible collinearity among some variables; for example,
basis~6 (restricted area 2-points) appears to be well approximated by a
combination of basis~4 (left restricted area 2-points) and basis~5 (right
restricted area 2-points).

\begin{table}
  \caption{Summaries of DIC and LPML for the models for Curry, Durant, Harden,
    and James with and without $\xi = 0$.}
  \label{tab:comp}
  \centering
  \begin{tabular}{c c rrrrr}
    \toprule
    & & & Curry & Durant & Harden & James\\
    \midrule
    Joint Model & DIC & $\xi \ne 0$ & 2391.3 & 2977.1 & 1744.4 & 760.8 \\
                     &     & $\xi = 0$ & 2379.2 & 2985.7 & 1753.7 & 802.0 \\
    & LPML & $\xi \ne 0$ & $-1195.7$ & $-1489.6$ & $-872.4$ & $-380.8$ \\
    &      & $\xi = 0$ & $-1189.6$ & $-1493.8$ & $-877.0$ & $-401.2$ \\[2ex]
    Intensity & DIC & $\xi \ne 0$ & 1352.8 & 1593.0 & 12.3 & $-1012.9$ \\
                    &     & $\xi = 0$   & 1352.5 & 1593.0 & 12.2 & $-1013.4$ \\
    & LPML & $\xi \ne 0$ & $-676.5$ & $-797.4$ & $-6.2$ & 506.3 \\
    &      & $\xi = 0$ & $-676.3$ & $-797.4$ & $-6.2$ & 506.5 \\[1ex]
    Mark & DIC & $\xi \ne 0$ & 1038.4 & 1384.1 & 1732.1 & 1773.7 \\
                     &     & $\xi = 0$ & 1026.7 & 1392.8 & 1741.5 & 1815.4\\
    & LPML & $\xi \ne 0$ & $-519.2$ & $-692.2$ & $-866.2$ & $-887.1$ \\
    &      & $\xi = 0$ & $-513.3$ & $-796.4$ & $-870.7$ & $-907.7$ \\
    \bottomrule
  \end{tabular}
\end{table}

Table~\ref{tab:comp} summarizes the DIC and LPML for the
full joint model and its two components. The smallest absolute difference is
8.6~in DIC and 4.2~in LPML for Durant; the largest absolute difference is 41.2
in DIC and 20.4 in LPML for James.
The DIC has a rule of thumb similar to AIC in decision making
\citep[Page. 613]{spiegelhalter2002bayesian}: a difference larger than~10 is
substantial and a difference about 2--3 does not give an evidence to
support one model over the other. For LPML, a
difference less than 0.5 is ``not worth more than to mention'' and
larger than 4.5 can be considered ``very strong'' \citep{kass1995bayes}.
With these guidelines applied to DIC and LPML, the mark model with
shot intensity included as a covariate has a clear advantage relative
to the model without it for Durant, Harden, and James, but not for Curry,
an interesting result which will be discussed in the next subsection.
The difference in DIC and LPML between the models with and without
$\xi = 0$ comes from the mark component. The two criteria for the intensity
component are almost the same with and without $\xi = 0$. This is expected
because the marks may contain little information about the
intensities, and intensity fitting results are not influenced by the
mark model significantly.

In order to have a direct comparison of improvement of mark model by
using the preferred model, we calculate the mean squared error (MSE)
of fitted mark models with and without intensity as a covariate. The
preferred models for all four players, which are intensity independent
model for Curry and intensity dependent model for other three players,
can reduce the MSE by 2.7\%, 1.3\%, 2.0\%, and 7.0\%.

\subsection{Fitted Results}

\begin{table}
  \caption{
    Estimated coefficients in the joint models for Curry, Durant, Harden, and James.}
  \label{tab:Realdatafit}
  \centering
  \begin{tabular}{cllrrr}
    \toprule
    & & & Posterior & Posterior & 95\% Credible\\
  Player & Model & Covariates & Mean & SD & Interval\\
    \midrule
Curry  & Intensity & baseline ($\lambda_0$)      & 0.236     & 0.012 & (\phantom{$-$}0.213, \phantom{$-$}0.261) \\
       &           & basis 1 (long 2-pointers)  & 0.248     & 0.041 & (\phantom{$-$}0.167, \phantom{$-$}0.328) \\ 
       &           & basis 2 (right wing threes)  & 0.290     & 0.025 & (\phantom{$-$}0.236, \phantom{$-$}0.335) \\ 
       &           & basis 3 (left wing threes)   & 0.190     & 0.028 & (\phantom{$-$}0.132, \phantom{$-$}0.243) \\ 
       &           & basis 4 (left restricted area) & 0.185     & 0.017 & (\phantom{$-$}0.152, \phantom{$-$}0.217) \\ 
       &           & basis 7 (top of key threes)  & 0.141     & 0.026 & (\phantom{$-$}0.091, \phantom{$-$}0.193) \\ 
       &           & basis 8 (center threes)     & 0.636     & 0.037 & (\phantom{$-$}0.563, \phantom{$-$}0.708) \\ 
       &           & basis 9 (corner threes)      & 0.121     & 0.019 & (\phantom{$-$}0.085, \phantom{$-$}0.158) \\ 
       & Mark      & intercept                   & $-0.165$  & 0.092 & ($-0.338$, \phantom{$-$}0.022) \\
       &           & distance                    & $-0.270$  & 0.064 & ($-0.396$, $-0.145$) \\
Durant & Intensity & baseline ($\lambda_0$)      & 0.372     & 0.015 & (\phantom{$-$}0.342, \phantom{$-$}0.401) \\
       &           & basis 1 (long 2-pointers)  & 0.465     & 0.039 & (\phantom{$-$}0.386, \phantom{$-$}0.539) \\ 
       &           & basis 2 (right wing threes)  & 0.219     & 0.028 & (\phantom{$-$}0.163, \phantom{$-$}0.270) \\ 
       &           & basis 3 (left wing threes)   & 0.097     & 0.032 & (\phantom{$-$}0.036, \phantom{$-$}0.162) \\
       &           & basis 4 (left restricted area) & 0.149     & 0.028 & (\phantom{$-$}0.096, \phantom{$-$}0.206) \\
       &           & basis 6 (restricted area)  & $-0.107$  & 0.027 & ($-0.160$, $-0.056$) \\
       &           & basis 7 (top of key threes)  & 0.071     & 0.027 & (\phantom{$-$}0.014, \phantom{$-$}0.121) \\
       &           & basis 8 (center threes)   & 0.634     & 0.038 & (\phantom{$-$}0.562, \phantom{$-$}0.707) \\
       &           & basis 9 (corner threes)   & $-0.074$  & 0.036 & ($-0.147$, $-0.007$) \\
       &           & basis 10 (mid-range twos) & 0.479     & 0.036 & (\phantom{$-$}0.408, \phantom{$-$}0.550) \\
       & Mark      & intercept                   & $-0.353$  & 0.127 & ($-0.609$, $-0.114$) \\ 
       &           & intensity ($\lambda$)       & 1.237     & 0.430 & (\phantom{$-$}0.393, \phantom{$-$}2.065) \\ 
       &           & distance                    & $-0.351$  & 0.068 & ($-0.481$, $-0.209$) \\
Harden & Intensity & baseline ($\lambda_0$)      & 0.348     & 0.015 & (\phantom{$-$}0.319, \phantom{$-$}0.378) \\
       &           & basis 1 (long 2-pointers)  & $-0.169$  & 0.045 & ($-0.258$, $-0.085$) \\ 
       &           & basis 2 (right wing threes)  & 0.193     & 0.021 & (\phantom{$-$}0.154, \phantom{$-$}0.236) \\ 
       &           & basis 3 (left wing threes)   & 0.084     & 0.025 & (\phantom{$-$}0.038, \phantom{$-$}0.135) \\ 
       &           & basis 4 (left restricted area) & 0.235     & 0.023 & (\phantom{$-$}0.186, \phantom{$-$}0.277) \\
       &           & basis 6 (restricted area)  & 0.127     & 0.022 & (\phantom{$-$}0.085, \phantom{$-$}0.172) \\
       &           & basis 7 (top of key threes)  & 0.247     & 0.018 & (\phantom{$-$}0.209, \phantom{$-$}0.281) \\
       &           & basis 8 (center threes)    & 0.657     & 0.029 & (\phantom{$-$}0.598, \phantom{$-$}0.712) \\
       &           & basis 10 (mid-range twos) & 0.086     & 0.023 & (\phantom{$-$}0.043, \phantom{$-$}0.133) \\
       & Mark      & intercept                   & $-0.453$  & 0.067 & ($-0.582$, $-0.323$) \\
       &           & intensity ($\lambda$)       & 1.291     & 0.201 & (\phantom{$-$}0.903, \phantom{$-$}1.686) \\ 
James & Intensity & baseline ($\lambda_0$)       & 0.423     & 0.016  & (\phantom{$-$}0.395, \phantom{$-$}0.457) \\
      &           & basis 1 (long 2-pointers)   & 0.113    & 0.035  & (\phantom{$-$}0.045, \phantom{$-$}0.181) \\ 
      &           & basis 3 (left wing threes)    & 0.165    & 0.028  & (\phantom{$-$}0.114, \phantom{$-$}0.223) \\
      &           & basis 4 (left restricted area) & 0.166    & 0.019  & (\phantom{$-$}0.128, \phantom{$-$}0.204) \\
      &           & basis 6 (restricted area)   & 0.130    & 0.016  & (\phantom{$-$}0.098, \phantom{$-$}0.161) \\
      &           & basis 7 (top of key threes)   & 0.087    & 0.026  & (\phantom{$-$}0.037, \phantom{$-$}0.136) \\
      &           & basis 8 (center threes)        & 0.544    & 0.031  & (\phantom{$-$}0.483, \phantom{$-$}0.603) \\
      &           & basis 9 (corner threes)       & 0.069    & 0.023  & (\phantom{$-$}0.023, \phantom{$-$}0.111) \\
      &           & basis 10 (mid-range twos) & 0.246    & 0.025  & (\phantom{$-$}0.200, \phantom{$-$}0.296) \\
      & Mark      & intercept                    & $-0.447$ & 0.073  & ($-0.588$, $-0.304$) \\ 
      &           & intensity ($\lambda$)        & 0.632    & 0.115 & (\phantom{$-$}0.418, \phantom{$-$}0.861) \\ 
      &           & distance                     & $-0.326$ & 0.056 & ($-0.433$, $-0.207$) \\ 
    \bottomrule
  \end{tabular}
\end{table}

Table~\ref{tab:Realdatafit} summarizes the posterior mean, posterior standard
deviation, and the 95\% highest posterior density (HPD) credible intervals for
the regression coefficients in the models for Curry, Durant, Harden, and James
as selected by the DIC and LPML.
Only significant covariates are displayed as determined by whether or not the
95\% HPD credible intervals cover zero in the first run. The reported results were
from the second run after insignificant covariates were removed.

\begin{figure}[tbp]
\centering
Fitted Intensity Surface 
\includegraphics[width = \textwidth]{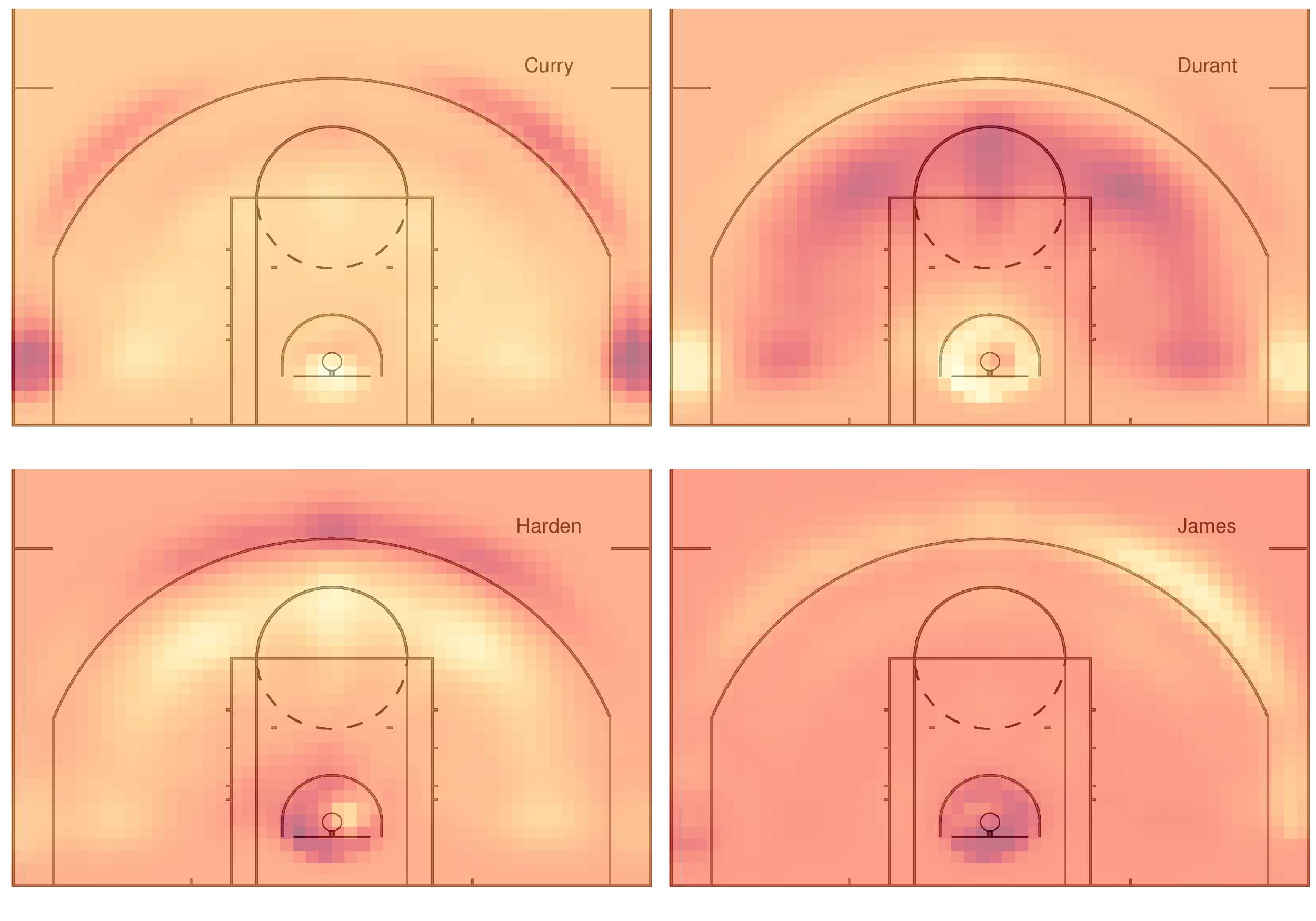}
Fitted Score Surface 
\includegraphics[width = \textwidth]{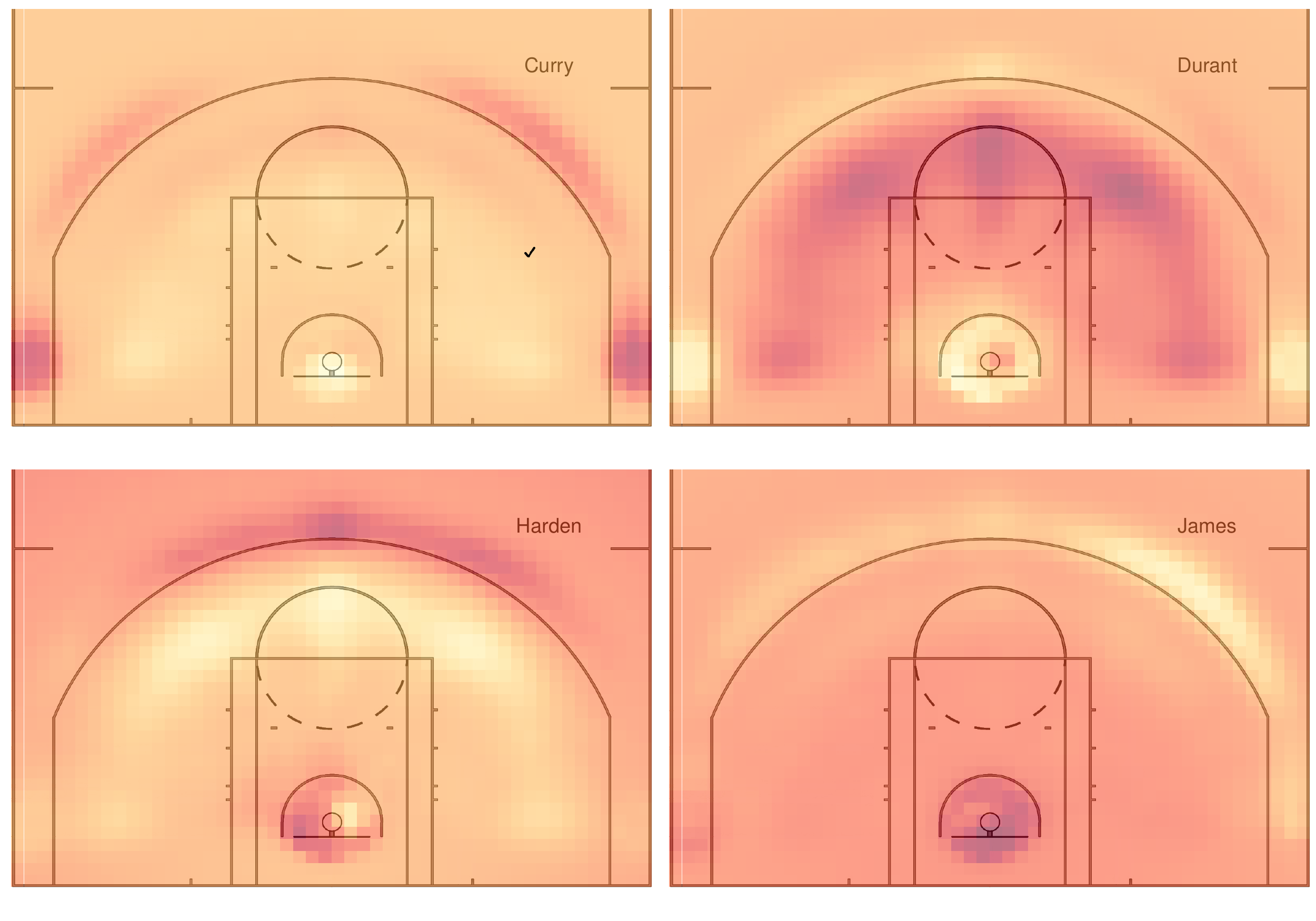}
\caption{Fitted shot intensity surfaces (upper) and expected score surfaces
  (lower) of Curry, Durant, Harden and James on
  the same scale. Redder means higher.}
\label{fig:heatmap}
\end{figure}

The coefficients of the 10 basis shot styles in the intensity model describe the
composition of each individual player's shot style. After being exponentiated,
they represent a multiplicative effect on the baseline intensity. So they are
comparable across players as a relative scale. The four players are quite
different in the coefficients of a few well-interpreted bases. Curry's rate of
corner threes was the highest among the four. Durant has the least rate
of corner threes and highest rate of long/mid-range 2-pointers. Harden had the
least rate of long 2-pointers and highest rate of top of key threes. Curry and
Harden had less two point shots but more three point shots than Durant and
James. James seemed to prefer to shot on the left side of court for three point
shots more than the other three players. All four had high rate of center threes.
Figure~\ref{fig:heatmap} (upper) shows the fitted intensity surfaces of the
four players. These results echo the findings in earlier works
\citep{miller2014factorized, franks2015characterizing}.

The results from the mark model conditioning on the intensity are the major
contribution of this work. All non-spatial covariates were
insignificant and were dropped from the model, except shot distance.
The coefficient of the intensity was found to be significantly positive for
Durant, Harden, and James, but not for Curry. That is, for the players
excluding Curry,  shot accuracy was higher where they shot more frequently.
The interaction between the intensity of shot type (2- vs 3-point) was not
significant for any player, suggesting that, for those whose shot frequency
and shot accuracy were positively associated, the association was not influenced
directly by shot rewards. The magnitude of coefficient of the intensity shows
how strong this dependence is. The association is much weaker (about a half) for
James compared to Durant and Harden. Shot distance was found to have a
significantly negative effect on shot accuracy for Curry, Durant, and James,
but not for Harden. The presence of both shot distance and intensity in the shot
accuracy model means that among locations with the same accuracy but
different rewards (2-
vs 3-point), 3-point locations tend to have higher intensities. This reflects
the bias of shooting intensity to 3-point shot due to higher rewards
\citep{alferink2009generality}. Since shot distance was not significant in
Harden's model, he could make more 3-point shots for higher rewards.

Curry's mark model only included a single covariate shot distance with a
significantly negative coefficient. At shot locations with the same shot
distance, Curry's shot accuracy was not affected by his shot frequency, which
makes him hard to guard against for a defense team. From an alternative
direction of reasoning, Curry's results suggest that he did not shoot more often
at locations where his shot accuracy was higher, which might not be optimal
from the team strategy point of view. The might be due to his injury in
that season and reduced time on court. He could make more shots where his
accuracy is higher to improve scoring efficiency. 

The fitted mark model allows combining shot accuracy and shot frequency to
construct an expected score map for each player; see Figure~\ref{fig:heatmap}
(lower). This plot is more informative than a shooting accuracy plot because the
latter would contain no value at locations where there were few or no
shots. Curry had a more  obvious scoring pattern of corner threes among the four.
Durant and James had more two point scores and less three point scores than
Curry and Harden. Curry and Harden's two point scores were more concentrated
in the restricted area than Durant and James.

To get an idea about the intensity dependent effect on shot accuracy averaged
over top players, we analyzed all shots attempted by the top 20 most frequent
shooters, which cover Harden and James, but not Curry and Durant. The 20 players'
data were pooled and treated as one virtual player.
Due to computational feasibility, we could not include more players in the pool.
The fitted coefficient of the intensity divided by 20
gives an ``elite average''  of the intensity dependent effect, which is
1.023. Compared with the results in Table~\ref{tab:Realdatafit}, Harden and Durant's
fitted coefficients were above the average, while James and Curry's were below the
average (Curry's fitted coefficient can be treated as 0 since his result favors
intensity independent model). The ranking relative to the elite average could be
a measure in assessing the players' efficiency in shot location
selection. Players with a fitted coefficient below the elite average might have
room to improve their score efficiency through shot selection.

\subsection{Application to Top 50 Most Frequent Shooters}

We further applied the same analysis to each of the top 50 most frequent
shooters in the 2017--2018 regular season. The number of shots of the
50 players ranged from 813 (Andre Drummond) to 1,517 (Russell Westbrook).
Among them, 40 players' data favored the intensity dependent model ($\xi \ne 0$)
in terms of DIC and LPML. Their estimated coefficients of the intensity in the
mark model were all positive; the interactions between the intensity and the shot
type (2- vs 3-point) were all not significantly different from zero. That is, 80\% of the
most frequent shooters in that season had positive association between shot
intensity and shot accuracy, and the association did not vary with shot
rewards. For the 10 players who had intensity independent mark models
similar to Curry, shot distance was found to be significantly negative in every
model.

\begin{figure}[tbp]
  \centering
  \subfloat[][Intensity Model]{
    \includegraphics[width = 0.486\textwidth]{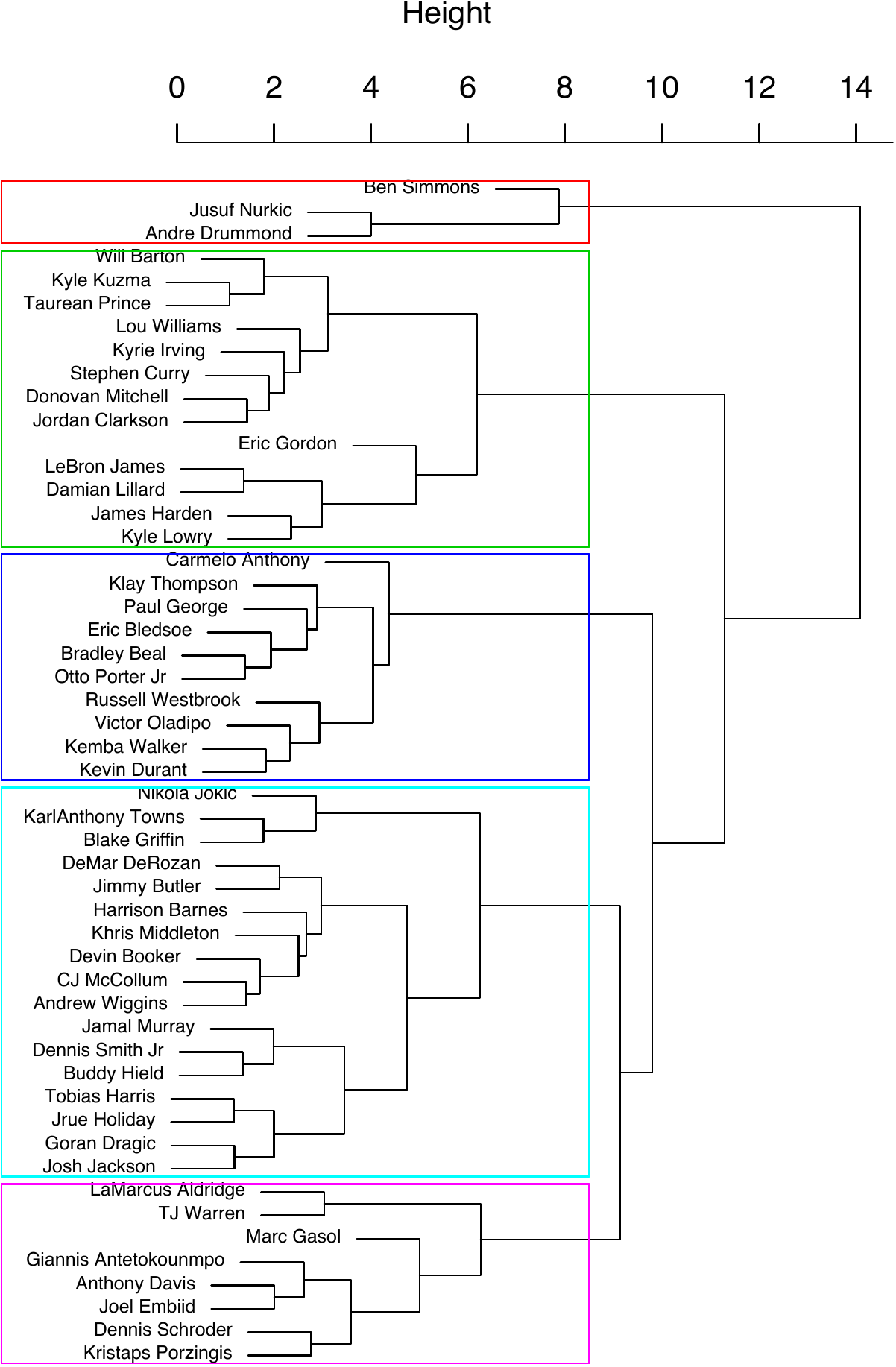}
    \label{fig:clus:inten}
  }
  \subfloat[][Mark Model]{
    \includegraphics[width = 0.486\textwidth]{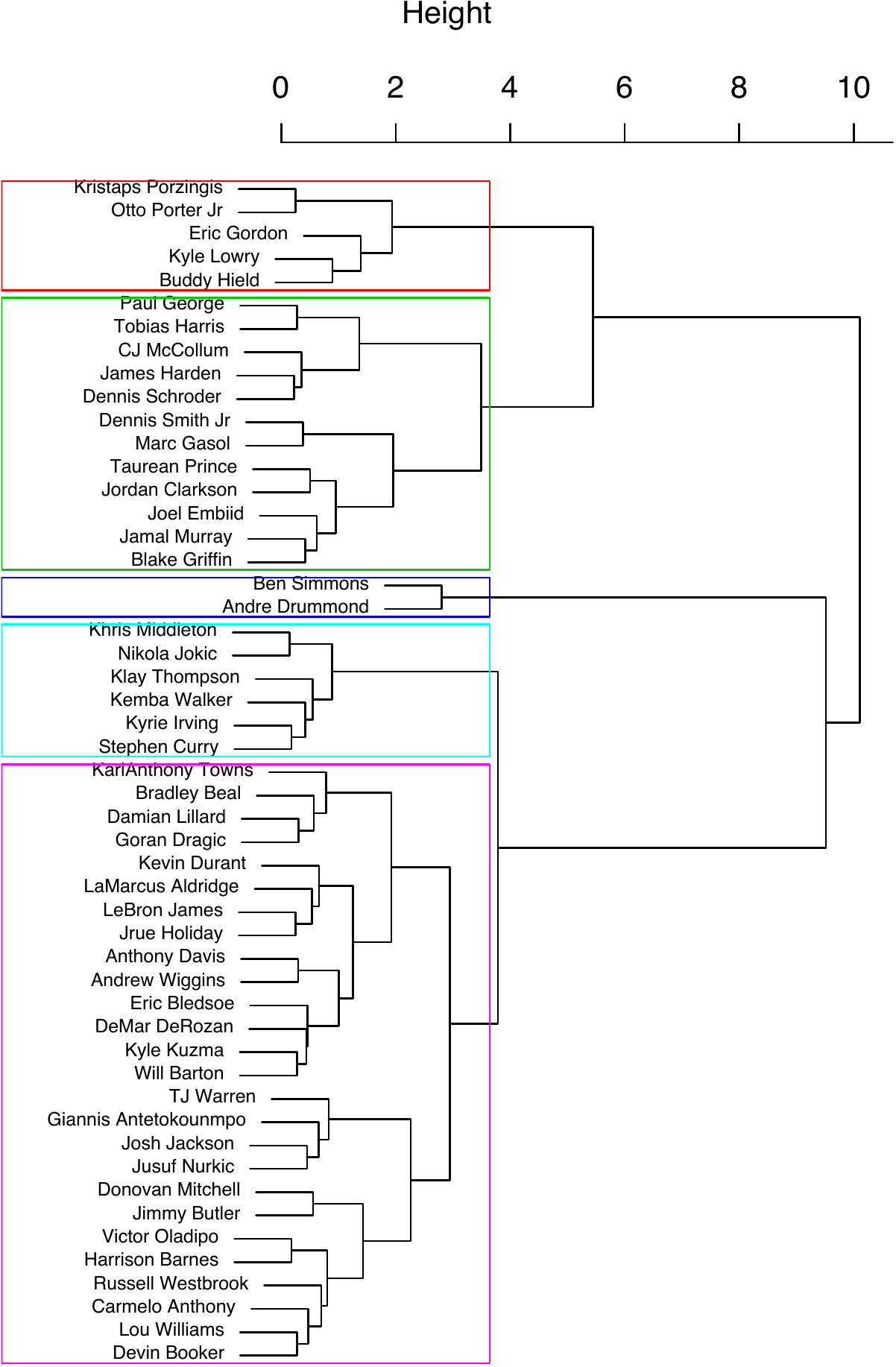}
    \label{fig:clus:mark}
  }
  \caption{Hierarchical clustering of 51 NBA players into 5 groups based on 
    fitted coefficients in the intensity model and the mark model.}
  \label{fig:cluster}
\end{figure}

The estimated coefficients in the joint model can be used as features to cluster
the players into groups. With Curry added in, estimates from a total of 51
players were used as inputs to a cluster analysis. Both clustering for shot
patterns based on the estimated coefficients in the intensity model and
clustering for the accuracy-intensity relationship based on the mark model given
intensity were considered. For the shot pattern clustering, only 10 coefficients
of the basis styles, with the baseline intensity excluded, were used to focus on
the distribution of the pattern instead of the total count of shots. The
clustering of the accuracy-intensity relationship clustering only used the
coefficients of intensity and shot distance in addition to the intercept because
the other coefficients were found to be insignificant for most of the players.
We used the hierarchical clustering method using the minimum variance criterion
of \citet{ward1963hierarchical} as implemented in R \citep{murtagh2014ward}.

Figure~\ref{fig:clus:inten} displays the results of clustering the 51 players by
their shot patterns into 5 groups. The first group only contains three
players who made mostly 2-point shots. The second group includes, interestingly,
Curry, Harden, and James. The closest players to Curry, Durant, and James were,
respectively, Kyrie Irving, Kyle Lowry, and Damian Lillard. Players in this
group had relative small coefficients for bases~1 and~10, and large coefficients
for bases~3 and~8. That is, they had less long/mid-range twos and more threes,
especially left wing threes. Players in the third group, which includes Durant, had
large coefficients for bases~1 and~10, showing that they had more long/mid-range
2-pointers. The closest player to Durant was Kemba Walker. Group four includes
players with small coefficient for basis~6 and large coefficient for basis~9,
which means that they had less 2-pointers from the restricted area and more
corner threes. The last group contains to players with small coefficients for
bases~3 and~9, and large coefficient for basis 10, indicating less left wing
threes and corner threes, but more mid-range twos.

The clustering results of the 51 players by the characteristics of their shot
accuracy in relation to their shot intensity are shown in Figure~\ref{fig:clus:mark}.
Group two has Harden and other players whose mark model contained the shot
intensity but not distance. Group four, which includes Curry, contains half of the
players whose shot intensity was insignificant in their mark model. Group five
is the largest group, which includes Durant and James. The players in this
group had significant shot distance effect on their accuracy. Most of them had
intensity in the mark model with a relatively small coefficients, and five of them
had intensity insignificant. The first group includes players with intensity but
not distance in the mark model, which is similar to Group~two, but the magnitude
of the coefficient for the intensity was the largest among all the players,
suggesting the strongest dependence between shot intensity and shot accuracy.
Players here were more likely to shoot at locations with higher accuracy
rates. The third group has only two players, Simmons and Drummond, whose
coefficients for shot distance were much larger than others' in magnitude, which
was expected because the two players shot mostly in the restricted area.

\section{Discussion}\label{sec:disc}

We proposed a Bayesian marked spatial point process to model both the shot
locations and shot outcomes in NBA players' shot charts. Basis shot styles
constructed from the NMF method \citep{miller2014factorized} were included as
covariates in the intensity for the Poisson point process model and the logistic
model for shot outcomes. For a majority of the top players, a positive
association between the shot intensity and shot accuracy was reported. The
association did not vary significantly according to the shot rewards. Players
whose shot intensity was not found to affect their shot accuracy (e.g., Curry)
may be hard to defend against. From the offense perspective, these players could
score more by making more shots where they shot more frequently. 
The cluster analyses based on the fitted coefficients characterizing the shot
pattern and shot accuracy are quite unique. Unlike other cluster analyses,
\citep[e.g.,][]{zhang2018clustering}, the data input here are not directly
observed but estimated from fitting a model to the shot charts. Consequently,
less obvious insights could be discovered.

A few directions of further work are worth investigating. 
Our proposed model is univariate in the sense that each player is modeled
separately. A full hierarchical model for pooled data from multiple players in
one season may be useful with a random effect at the player level for certain
parameters. The number of basis shot styles was set to 10 as suggested by
\citet{miller2014factorized}. It would be interesting to find an optimal number
of basis through model comparison criteria like DIC and LPML. An important
factor for shot accuracy is the shot clock time remaining
\citep{skinner2012problem}, but it is not available in the dataset we
obtained. It should be added to the mark model if available. Our
spatial Poisson process model formulates a linear relationship between
the spatial covariates and the log intensity, which cannot capture more
complicated spatial trend of the intensity of spatial point pattern. Including
some Bayesian non-parametric methods like finite mixture model
\citep{miller2018mixture} may help increase the
accuracy of the estimation of spatial point pattern.

\section*{Appendix}
This section shows the tables of simulation results.

\begin{table}
  \caption{Summaries of the bias, standard deviation (SD), average of the
    Bayesian SD estimate ($\widehat{\mathrm{SD}}$), and coverage rate
    (CR) of 95\% credible intervals when $Z_2$ is continuous:
    $\xi = \alpha_0 = 0.5$, $\alpha_2 = 1$, $(\beta_1, \beta_2) = (2,
    1)$ and $Z_2 \sim N(0, 1)$.}
  \label{tab:EstSimu_norm}
  \centering
  \begin{tabular}{crrrrrrrrrr}
    \toprule
     & & & \multicolumn{4}{c}{$\lambda_0 = 0.5$} &
       \multicolumn{4}{c}{$\lambda_0 = 1$}\\
    \cmidrule(lr){4-7} \cmidrule(lr){8-11}
    $\alpha_1$ &Model & Para & Bias & SD & $\widehat{\mathrm{SD}}$ & CR
& Bias &SD & $\widehat{\mathrm{SD}}$ & CR\\
    \midrule
    0.8 & Intensity & $\lambda_0$ & 0.01 & 0.04 & 0.04 & 0.96 & 0.01 & 0.06 & 0.06 & 0.93 \\ 
        &           & $\beta_1$   & $-0.06$ & 0.11 & 0.11 & 0.90 & $-0.06$ & 0.09 & 0.08 & 0.88 \\ 
        &           & $\beta_2$ & $-0.05$ & 0.09 & 0.09 & 0.94 & $-0.03$ & 0.06 & 0.06 & 0.92 \\ 
        & Mark      & $\xi$ & 0.11 & 0.57 & 0.60 & 0.97 & 0.04 & 0.22 & 0.22 & 0.96 \\ 
        &           & $\alpha_0$ & 0.01 & 0.20 & 0.20 & 0.95 & 0.00 & 0.14 & 0.14 & 0.97 \\ 
        &           & $\alpha_1$ & 0.03 & 0.13 & 0.13 & 0.94 & 0.01 & 0.09 & 0.09 & 0.94 \\ 
        &           & $\alpha_2$ & 0.03 & 0.14 & 0.14 & 0.95 & 0.01 & 0.10 & 0.10 & 0.93 \\ 
\\
     1  & Intensity & $\lambda_0$ & 0.00 & 0.04 & 0.04 & 0.94 & 0.00 & 0.06 & 0.06 & 0.94 \\ 
        &           & $\beta_1$ & $-0.05$ & 0.11 & 0.11 & 0.94 & $-0.05$ & 0.08 & 0.08 & 0.91 \\ 
        &           & $\beta_2$ & $-0.03$ & 0.10 & 0.09 & 0.92 & $-0.04$ & 0.07 & 0.06 & 0.92 \\ 
        & Mark      & $\xi$ & 0.03 & 0.60 & 0.61 & 0.95 & 0.05 & 0.21 & 0.22 & 0.96 \\ 
        &           & $\alpha_0$ & 0.00 & 0.20 & 0.20 & 0.95 & $-0.01$ & 0.14 & 0.14 & 0.95 \\ 
        &           & $\alpha_1$ & 0.01 & 0.13 & 0.14 & 0.97 & 0.01 & 0.10 & 0.10 & 0.96 \\ 
        &           & $\alpha_2$ & 0.03 & 0.14 & 0.14 & 0.95 & 0.01 & 0.10 & 0.10 & 0.94 \\ 
\\
     2 & Intensity & $\lambda_0$ & 0.00 & 0.04 & 0.04 & 0.94 & 0.00 & 0.06 & 0.06 & 0.95 \\ 
       &           & $\beta_1$ & $-0.06$ & 0.12 & 0.11 & 0.91 & $-0.04$ & 0.08 & 0.08 & 0.93 \\ 
       &           & $\beta_2$ & $-0.03$ & 0.09 & 0.09 & 0.94 & $-0.03$ & 0.07 & 0.06 & 0.91 \\ 
       & Mark      & $\xi$ & 0.04 & 0.71 & 0.69 & 0.94 & 0.05 & 0.23 & 0.24 & 0.96 \\ 
       &           & $\alpha_0$ & 0.02 & 0.22 & 0.23 & 0.95 & $-0.01$ & 0.15 & 0.16 & 0.97 \\ 
       &           & $\alpha_1$ & 0.08 & 0.23 & 0.21 & 0.93 & 0.03 & 0.15 & 0.15 & 0.94 \\ 
       &           & $\alpha_2$ & 0.03 & 0.17 & 0.16 & 0.93 & 0.02 & 0.11 & 0.11 & 0.95 \\
    \bottomrule
  \end{tabular}
\end{table}

\begin{table}
  \centering
  \caption{Summaries of the bias, standard
    deviation (SD), average of the
    Bayesian SD estimate ($\widehat{\mathrm{SD}}$), and coverage rate
    (CR) of 95\% credible intervals when $Z_2$ is binary:
    $\xi = \alpha_0 = 0.5$, $\alpha_2 = 1$, $(\beta_1, \beta_2) = (2,
    1)$ and $Z_2 \sim Bernoulli(0.5)$.
  }
  \label{tab:EstSimu_binary}
  \begin{tabular}{crrrrrrrrrr}
    \toprule
     & & & \multicolumn{4}{c}{$\lambda_0 = 0.5$} &
       \multicolumn{4}{c}{$\lambda_0 = 1$}\\
    \cmidrule(lr){4-7} \cmidrule(lr){8-11}
    $\alpha_1$ &Model & Para & Bias & SD & $\widehat{\mathrm{SD}}$ & CR
& Bias &SD & $\widehat{\mathrm{SD}}$ & CR\\
    \midrule
    0.8 & Intensity & $\lambda_0$ & 0.00 & 0.04 & 0.04 & 0.94 & 0.01 & 0.06 & 0.06 & 0.95 \\ 
    && $\beta_1$ & $-0.05$ & 0.12 & 0.11 & 0.88 & $-0.05$ & 0.08 & 0.08 & 0.90 \\ 
    && $\beta_2$ & $-0.02$ & 0.10 & 0.09 & 0.94 & $-0.03$ & 0.06 & 0.06 & 0.95 \\ 
    & Mark & $\xi$ & 0.07 & 0.62 & 0.61 & 0.94 & 0.04 & 0.20 & 0.23 & 0.96 \\ 
    && $\alpha_0$ & 0.00 & 0.23 & 0.22 & 0.93 & 0.01 & 0.16 & 0.16 & 0.95 \\ 
    && $\alpha_1$ & 0.03 & 0.13 & 0.13 & 0.96 & 0.01 & 0.10 & 0.10 & 0.94 \\ 
    && $\alpha_2$ & 0.03 & 0.25 & 0.24 & 0.96 & 0.01 & 0.19 & 0.18 & 0.95 \\ 
    \\
    1 & Intensity & $\lambda_0$ & 0.00 & 0.04 & 0.04 & 0.94 & 0.00 & 0.06 & 0.06 & 0.94 \\ 
    && $\beta_1$ & $-0.06$ & 0.11 & 0.11 & 0.93 & $-0.04$ & 0.09 & 0.08 & 0.89 \\ 
    && $\beta_2$ & $-0.04$ & 0.08 & 0.09 & 0.94 & $-0.02$ & 0.06 & 0.06 & 0.93 \\ 
    & Mark & $\xi$ & 0.10 & 0.64 & 0.63 & 0.94 & 0.09 & 0.22 & 0.23 & 0.94 \\ 
    && $\alpha_0$ & 0.01 & 0.23 & 0.23 & 0.97 & $-0.03$ & 0.16 & 0.16 & 0.94 \\ 
    && $\alpha_1$ & 0.03 & 0.15 & 0.14 & 0.92 & 0.01 & 0.11 & 0.10 & 0.92 \\ 
    && $\alpha_2$ & 0.02 & 0.27 & 0.25 & 0.93 & 0.02 & 0.17 & 0.18 & 0.96 \\ 
    \\
    2 & Intensity & $\lambda_0$ & 0.00 & 0.04 & 0.04 & 0.95 & 0.01 & 0.06 & 0.06 & 0.94 \\ 
    && $\beta_1$ & $-0.05$ & 0.11 & 0.11 & 0.92 & $-0.05$ & 0.08 & 0.08 & 0.90 \\ 
    && $\beta_2$ & $-0.04$ & 0.09 & 0.09 & 0.91 & $-0.04$ & 0.06 & 0.06 & 0.92 \\ 
    & Mark & $\xi$ & 0.06 & 0.73 & 0.70 & 0.94 & 0.07 & 0.28 & 0.25 & 0.93 \\ 
    && $\alpha_0$ & 0.03 & 0.29 & 0.26 & 0.93 & $-0.01$ & 0.20 & 0.19 & 0.94 \\ 
    && $\alpha_1$ & 0.06 & 0.21 & 0.21 & 0.94 & 0.05 & 0.15 & 0.15 & 0.94 \\ 
    && $\alpha_2$ & 0.03 & 0.31 & 0.28 & 0.93 & 0.03 & 0.19 & 0.20 & 0.94 \\
    \bottomrule
  \end{tabular}
\end{table}


\newpage

\bibliographystyle{chicago}
\bibliography{main}
\end{document}